# Crystalline β-$Ga_2O_3$ thin films deposited via reactive magnetron sputtering of a liquid Ga target

Running title: β-$Ga_2O_3$ thin films by reactive sputtering

Running Authors: Novák et al.

Petr Novák[1,a], Jan Koloros[2], Stanislav Haviar[2], Jiří Rezek[2], Pavel Baroch[2]

[1] New Technologies – Research Centre, University of West Bohemia in Pilsen, Pilsen, Czech Republic
[2] Department of Physics and NTIS - European Centre of Excellence, University of West Bohemia in Pilsen, Univerzitní 8, 301 00 Pilsen, Czech Republic

[a] Electronic mail: petrnov@ntc.zcu.cz

$Ga_2O_3$ thin films were deposited by reactive magnetron sputtering from a liquid gallium target at a fixed oxygen partial pressure of 0.1 Pa. The influence of deposition temperature, substrate type, and discharge parameters on the structural and electrical properties was systematically investigated. Films deposited on silicon and quartz glass exhibit polycrystalline growth, whereas sapphire substrates enable highly oriented growth of β-$Ga_2O_3$ with a preferred (−201) orientation. The lowest electrical resistivity of $7 \times 10^3$ Ω·cm was obtained for films deposited on sapphire at a temperature of 585 °C. At this temperature, the films reach sufficient crystalline quality to enable efficient charge carrier transport and thus the manifestation of unintentional conductivity. At higher deposition temperatures, pronounced crystallization occurs; however, it is not homogeneous throughout the entire film thickness, which leads to a deterioration of the electrical properties. These results demonstrate that, despite intrinsic limitations, reactive magnetron sputtering can be successfully employed for the preparation of $Ga_2O_3$ thin films with optimized electrical properties when appropriate substrates, oxygen partial pressure and deposition temperatures are selected.

# I. INTRODUCTION

Gallium oxide ($Ga_2O_3$) is a technologically important semiconductor material that has gained increasing research interest in recent years. Material performance exceeds known wide bandgap semiconductors such as GaN (3.4 eV) and SiC (3.3 eV). $Ga_2O_3$-based devices benefit from low power dissipation, high current density, high breakdown voltage and stable operation at high temperatures. Of the various phases of $Ga_2O_3$, the β-phase is the most stable at room temperature and pressure. It has a monoclinic crystal structure. All other $Ga_2O_3$ polymorphs transform into this phase upon heat treatment. β-$Ga_2O_3$ remains stable up to temperatures of around 1800 °C and is optically transparent in the deep ultraviolet (UV) region (below ~250 nm). Its electrical conductivity depends on growth conditions and defect chemistry, particularly oxygen vacancies. Due to its wide bandgap of approximately 4.7 eV, high thermal stability and chemical robustness, β-$Ga_2O_3$ is the primary phase of interest for use in power electronics, solar-blind UV photodetectors, gas sensors and transparent conducting films[1]. In addition, this material can be doped to exhibit n-type conductivity, further enhancing its utility in various electronic applications[2].

Given these unique properties, it is crucial to explore the new alternative growth methods that enable the development of β-$Ga_2O_3$, but each has its limitations[3]. Molecular beam epitaxy[4] offers precision and the ability to grow ultrathin films, but suffers from low growth rates and high operating costs. Metal-organic vapor phase epitaxy[5] offers higher growth rates and better control of material purity, but requires high temperatures that can affect film quality. Pulsed laser deposition[6] allows deposition at lower temperatures but has scalability issues and lower growth rates. Halide vapor phase epitaxy[7] achieves fast growth rates and produces high quality films, but suffers from rough surface finishes and requires further improvements in injector systems. Low-pressure chemical vapour deposition offers fast and scalable growth at relatively low equipment cost, but high deposition temperatures lead to element diffusion between films and substrates.

Despite the strengths of established epitaxial techniques, magnetron sputtering[8] offers advantages such as scalability and high deposition rates. However, it is currently less utilized for the preparation of high-quality $Ga_2O_3$ films due to the difficulty in achieving high crystallinity. Deposition from a ceramic $Ga_2O_3$ target occurs more frequently[9,10]. To the best of our knowledge, reactive deposition from a liquid target has been investigated only rarely[11], although similar deposition systems, such as those used for GaN[12,13] and $ZnGa_2O_4$ films[14], have been studied. This approach is enabled by the low melting point of gallium, which allows the film to be deposited from a liquid target. The fundamental processes occurring at the surface of the liquid target during reactive sputtering remain poorly understood and have not been described in detail. From a practical point of view, crystalline $Ga_2O_3$ films can already be prepared, with the substrate temperature around 600 °C being a key parameter[15,16]. Nevertheless, the deposited films are mainly evaluated in terms of their structural properties, while data on electrical properties, which are crucial for assessing suitability for typical $Ga_2O_3$ applications, are still largely missing. Although a limited number of studies report sputtered $Ga_2O_3$ films, electrical characterization is often neglected.

The objective of this study is to surmount these disadvantages identified and to produce $Ga_2O_3$ films of sufficient quality to study their electrical properties and enhance them for practical use. The preparation of films will be accomplished through the utilization of reactive pulsed magnetron sputtering, a method that affords a vast array of discharge parameter settings. This, in turn, will facilitate the optimization of the deposition process. The utilization of this method has traditionally been employed to achieve elevated deposition rates during reactive deposition. Nevertheless, the objective is to capitalize on the capacity to regulate the oxide conditions during the process, which have a profound effect on the electrical properties of the $Ga_2O_3$, due to the differing concentrations of defects that they engender. Initially, it is imperative to establish a $Ga_2O_3$ films that exhibits quantifiable electrical characteristics.

In this study, the growth behavior and crystallization of $Ga_2O_3$ films deposited by magnetron sputtering are investigated on different substrates. Non-epitaxial substrates (Si (100) and quartz glass) are used to identify the limitations of sputtering in the absence of epitaxial constraints, while sapphire, a typical epitaxial substrate for $Ga_2O_3$, enables an assessment of the relationship between deposition temperature, structural quality, and electrical resistivity. The study focuses on temperature-induced structural evolution as a key factor for achieving compact $Ga_2O_3$ films with reduced resistivity.

## II. EXPERIMENTAL

### A. Deposition

The $Ga_2O_3$ thin films were deposited using a pulsed reactive magnetron sputtering technique. A MAGPULS MP2-AS 400 power supply was used to drive an unbalanced magnetron sputtering system equipped with a liquid gallium target with a diameter of 100 mm and a thickness of 6 mm (target area At = 78.54 $cm^2$). The distance between the target and the substrate was approximately 10.5 cm. A scroll pump combined with a diffusion oil pump was used to evacuate the vacuum chamber to a base pressure of approximately $8 \times 10^{-4}$ Pa. One day prior to deposition, the vacuum chamber was preheated to 300 °C for 3 h and pre-pumped to improve the base pressure. The average applied power was varied between 50 and 200 W. The pulse length was set to 15 μs, while the off-time was varied between 15 and 985 μs in order to modulate peak power density delivered to the target. Depositions were performed on three types of substrates: monocrystalline silicon wafers with (100) orientation, quartz glass, and sapphire substrates with (0001) orientation. Prior to deposition, the substrates were cleaned sequentially in acetone, methanol, and distilled water for 10 min each and subsequently dried using nitrogen gas. The substrate temperature ranged from 520 °C to 660 °C. Film thicknesses of approximately 600 nm on Si (100) and approximately 900 nm on sapphire (0001) were obtained. The thickness on Si was measured using a profilometer, whereas the thickness on sapphire was determined from cross-sectional observations. The observed difference in film thickness is likely related to a slightly different substrate position relative to the erosion zone of the target. Under these gas conditions, a deposition rate of 15 nm $min^{-1}$ on silicon (100) was achieved: the argon partial pressure was maintained at 1.0 Pa and the oxygen partial pressure at 0.1 Pa to ensure reactive conditions suitable for the formation of $Ga_2O_3$.

### B. $Ga_2O_3$ films analysis

The electrical resistivity of the $Ga_2O_3$ films was measured using a standard four-point tungsten probe method in order to eliminate the influence of contact resistance. The measured square samples had dimensions of 10 × 10 mm², the distance between the probes was 1 mm, and the probe tip radius was 150 μm. Due to significant uncertainties in the carrier concentration and mobility values extracted from Hall effect measurements, only resistivity data are discussed in this work. The cross-sectional and surface morphology of the $Ga_2O_3$ films were examined using a scanning electron microscope (Hitachi SU-70) operated at acceleration voltages ranging from 2 to 5 kV. Film thickness was measured using a Dektak Pro profilometer. The crystal structure of the films was characterized by XRD diffraction (XRD) using a PANalytical X'Pert PRO diffractometer operating with Cu Kα radiation (λ = 0.154187 nm). Optical transmittance spectra were measured using an Agilent Technologies CARY 7000 spectrophotometer. The wavelength range used for the transmittance measurements was from 200 to 1500 nm.

## III. RESULTS AND DISCUSSION

The experimental findings are presented in a sequence that reflects the causal relationship between discharge conditions, film growth, and resulting structural and electrical properties. The initial focus of this study is on the examination of pulsed discharge parameters and oxidation conditions during reactive sputtering. The rationale behind this choice is that, given the established fact that the oxidation regime has a significant influence on the electrical behavior of the material, particularly with regard to the concentration of free charge carriers, this area is of particular relevance to the present study.

### A. Discharge characteristics and oxidation conditions

In order to investigate possible changes in the nature of the discharge characteristics (especially the I-V curve) for different target states, a series of current and voltage waveforms was measured for different average magnetron powers in the range of 20-620 W. (see Figure 1). For all experiments, the same $t_{on}/t_{off}$ = 15/45 μs and partial pressures of argon and oxygen were maintained at 1.0 Pa and 0.1 Pa, respectively. Figure 1 shows the classical dependence of the pulse voltage and current on the magnetron when the average magnetron power increases. The increasing average magnetron power, accompanied by concurrent increases in the target voltage (causing higher energy gain of secondary electrons in the target's sheath) and current (indicating an increase in the electron density during the voltage pulse), leads to more intensive ionization of the sputtered atoms[9].

Figure 2 shows qualitative changes of the target surface during the discharge. Even at low power levels, the target cannot be maintained in a solid state due to the low melting point of gallium of approximately 30 °C. At an average power of 40 W (or if $t_{off}$ = 985 μs), thin floating surface layers are observed on the surface of the molten target. With increasing power (or decreasing $t_{off}$), these surface layers gradually disappear. In the case of a liquid target, no typical erosion zone is formed and the target surface remains flat. In the present work, the chemical composition of the surface oxide was not directly analyzed. Under ambient conditions, liquid gallium is known to form an ultrathin (~5 Å) oxide layer commonly described as stoichiometric $Ga_2O_3$, typically amorphous or poorly crystalline[17]. However, under plasma conditions and ion bombardment, deviations from perfect stoichiometry are possible, and the surface layer may correspond to a defective

$Ga_2O_{3-x}$ or substoichiometric $GaO_x$ phase. A detailed determination of the exact chemical composition under the present discharge conditions would require dedicated surface-sensitive analysis and remains a subject for further investigation.

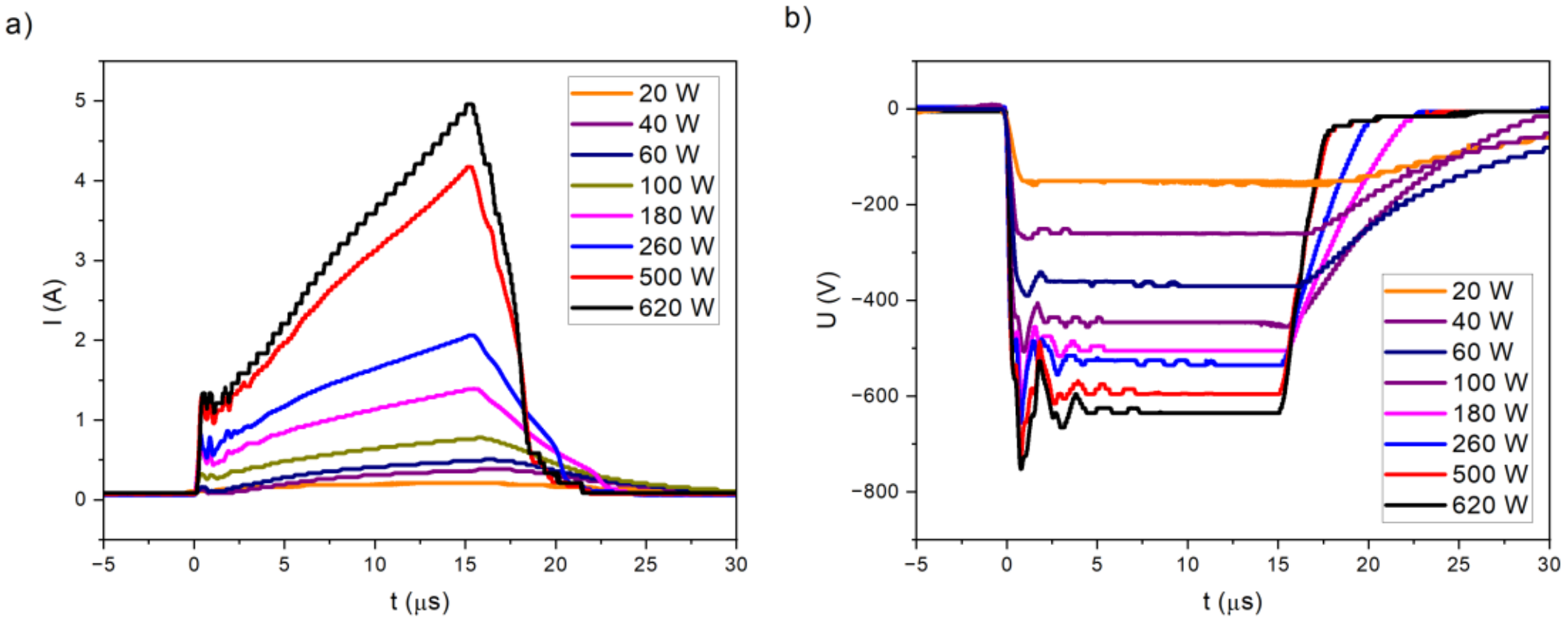

FIG. 1. Target current (a) and voltage (b) waveforms for different values of average power in pulse

FIG. 2. Picture of the Gallium target at (a) $t_{off}$ = 15 μs and (b) at the 985 μs.

The fundamental parameter of the discharge during our experiment is the off-time $t_{off}$, defined as the time interval between successive pulses when the power supply is at zero. Increasing this interval results in a higher energy density of the pulse, while the average power of the source remains constant. This has a direct impact on the nature of the discharge. Under short $t_{off}$ conditions, a movement of the target mass in the middle of the target is observed (speed increases with the average power and lowering the $t_{off}$). As the $t_{off}$ increases, oxide islands form and subsequently move along the target, coalescing at its edges and in the middle. In the last step, the layer in the center begins to connect with the layer on the edge at $t_{off}$ of 985 μs. Figure 3b shows that oxygen consumption decreases with increasing $t_{off}$. This corresponds to the expansion of the island on the surface of the target. It is analogous to target poisoning, which, in this case, is combined with the growth of an oxide phase in the form of a thin layer on the surface of the liquid gallium target.

The following chapter demonstrates that the XRD lines are most intense and narrowest within the $t_{off}$ range between 15 and 85 µs. This range is marked in orange. At an average power of 50 W, the $O_2$ flow rate remains nearly constant over the investigated range. In contrast, at higher power levels of 100 W and 150 W, a pronounced decrease in the required $O_2$ flow is observed. This behavior indicates a significant shift in the oxidation conditions with increasing average pulse energy, suggesting a strong coupling between the pulsed discharge parameters and the reactive gas consumption. The increase in discharge voltage with longer $t_{off}$ (Fig. 3a) is primarily driven by the energy density per pulse. Since the average magnetron power is maintained constant, increasing the off-time necessitates higher peak power and pulse voltage to compensate for the longer dead time. This electronic modulation effectively overrides the potential decrease in discharge voltage that would otherwise be expected from the higher secondary electron emission coefficient of the forming $Ga_2O_3$ layer. The qualitative increase in the secondary electron emission yield upon target oxidation was verified by comparing the discharge voltages in metallic and oxide modes, following the established methodology[18].

The $O_2$ flow rate (Fig. 3b) was automatically regulated to maintain a constant partial pressure of 0.1 Pa, and its behavior is directly linked to the "gettering effect" and the target's state. At short $t_{off}$ (15–135 µs), the high pulsing frequency effectively cleans the liquid gallium surface, maintaining a high deposition rate of approximately 15 nm/min (at 100 W). The resulting high flux of sputtered Ga atoms acts as an efficient getter, consuming a large amount of oxygen in the chamber, which necessitates a higher $O_2$ flow. In contrast, at $t_{off}$ = 985 µs, the target becomes poisoned by a continuous oxide layer, and the deposition rate drops significantly to 6 nm/min. With fewer Ga atoms available to react with the gas, the oxygen consumption decreases, and the automated system reduces the $O_2$ flow to maintain the set pressure.

An opposite influence of the pulse off-time on the oxidation conditions was reported previously for Al-doped ZnO thin films[19], where higher $t_{off}$ led to less oxidized deposits. This indicates that while the control of oxidation via pulse timing is a general feature, the specific response depends on the material system and requires further research focused on the discharge-surface kinetics.

Based on these parameters, we selected a power value of 100 W for film preparation. At this power value, changes in oxygen consumption can be observed, and it can be assumed that oxide conditions during deposition can be controlled simply by changing the $t_{off}$ parameter within the range of 15–85 μs.

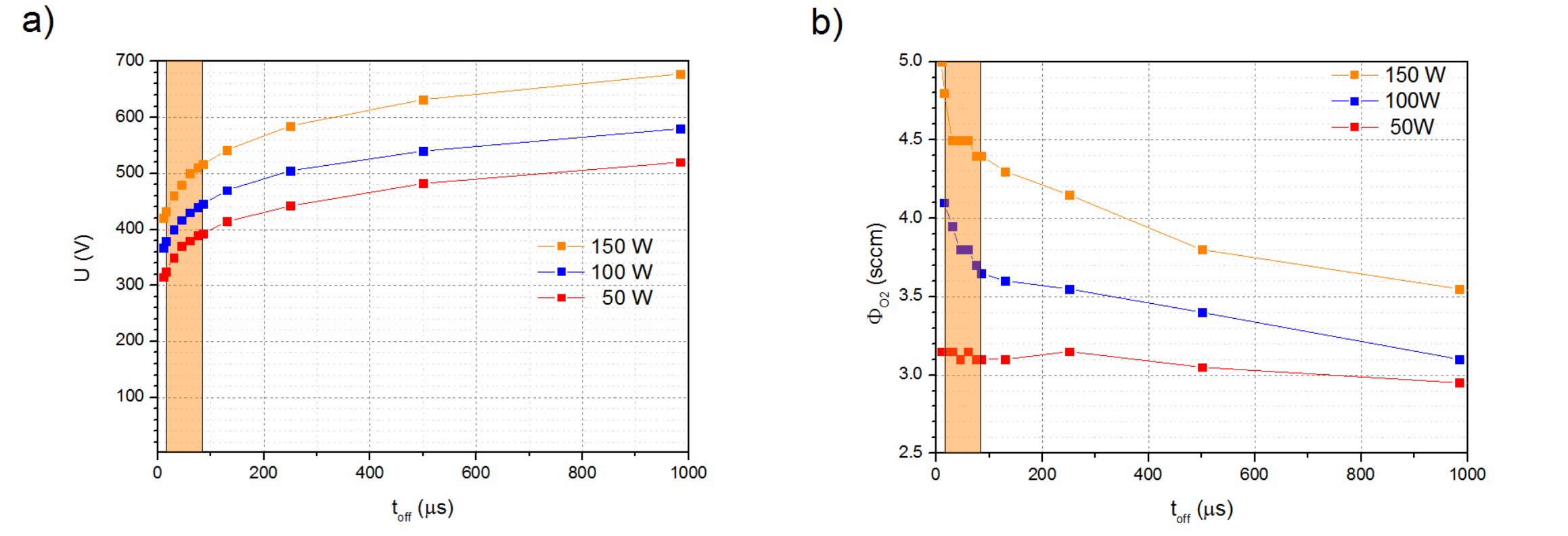

FIG. 3. Graphs show a) the magnetron discharge voltage as functions of $t_{off}$. And b) the oxygen flow rate $\Phi_{o2}$

-

### B. *Structural evolution of $Ga_2O_3$ films on non-epitaxial substrates (Si and quartz glass)*

The pulsed discharge parameters, particularly the off-time duration $t_{off}$, have a pronounced influence on the discharge characteristics and the effective energy input during film growth. In order to investigate the effect of $t_{off}$ on the resulting film structure, $Ga_2O_3$ films were deposited at a substrate temperature of 600 °C on silicon and $SiO_2$ substrates. These substrates were selected as non-epitaxial reference systems, thereby enabling the influence of discharge parameters on film crystallization to be evaluated independently of epitaxial effects. The structural properties of the deposited films were analyzed by XRD, and the corresponding diffractograms are presented in Figure 4. The film deposited on Si(100) exhibits a polycrystalline $\beta$-$Ga_2O_3$ structure, whereas the film grown on quartz shows a $\beta$-$Ga_2O_3$ phase with a dominant diffraction peak corresponding to the (400) plane. The XRD patterns demonstrate that films deposited within the $t_off$ range from 15 to 135 µs exhibit enhanced crystallinity, as evidenced by increased diffraction peak intensities and improved peak definition. For longer off-time durations, a degradation of the crystalline structure becomes evident.

This behavior can be attributed to an increased contribution of energetic particle bombardment, which disrupts the growth process and promotes defect formation within the film. Based on these observations, a process window for $t_{off}$ was identified, within which enhanced structural quality is achieved. This parameter range was subsequently utilized for further experiments, as a sufficient degree of crystallinity is a prerequisite for the subsequent investigation of electrical properties. This area is marked in Figure 3, and even at a power of 100 W, a significant change in oxygen consumption and discharge voltage is evident, which opens up possibilities for further optimizing discharge parameters.

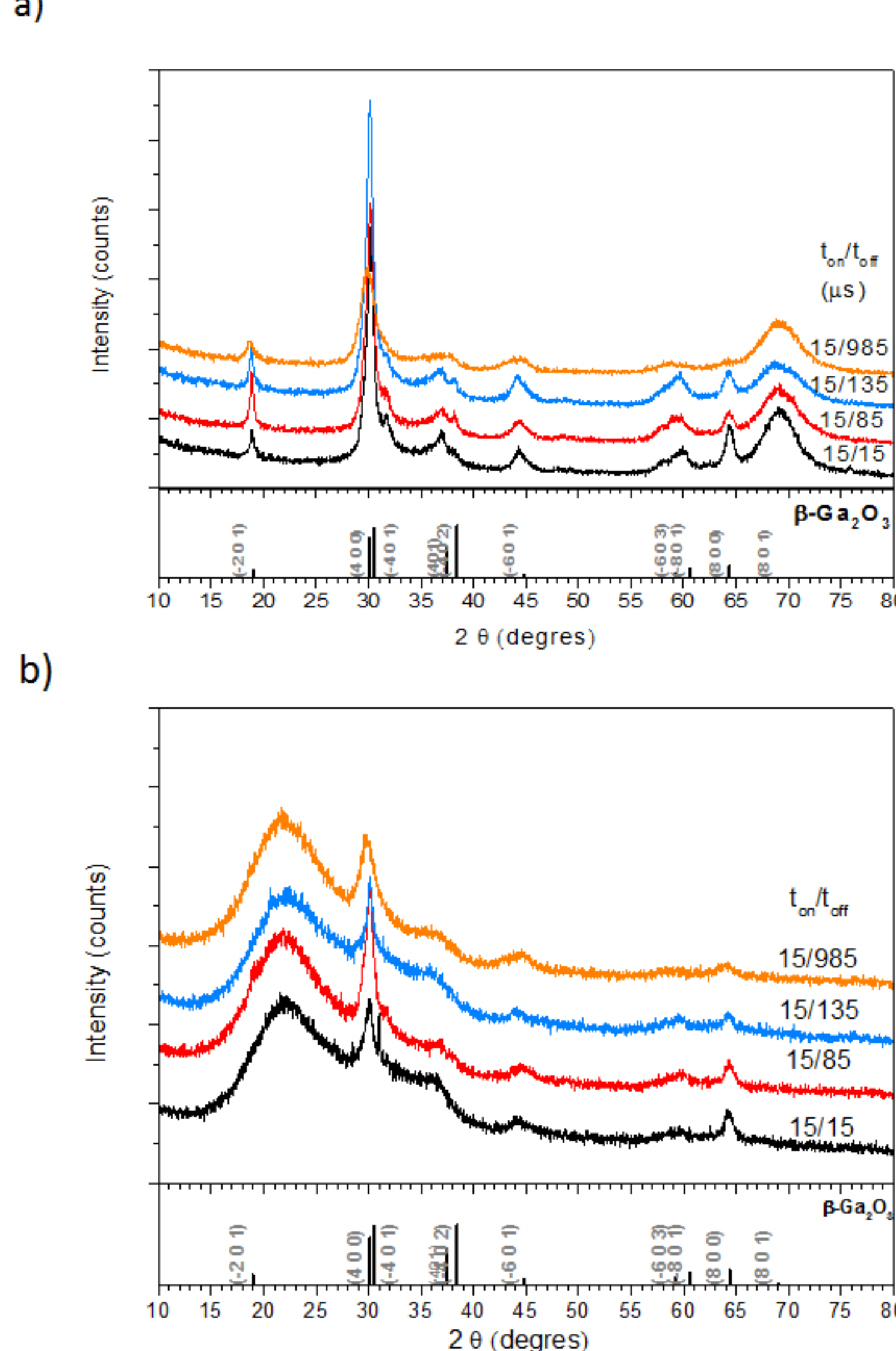

FIG. 4. XRD patterns of the films deposited at different $t_{off}$ on (a) Si (100) and (b) quartz glass. Grey vertical lines indicate selected prominent reflections of the $\beta$-$Ga_2O_3$ powder standard (PDF 01-087-1901), labeled by their corresponding (hkl) indices.

Within the identified $t_{off}$ window, the influence of substrate temperature on the crystallization behavior and structural evolution of the $Ga_2O_3$ films was subsequently investigated. Based on previous reports, substrate temperatures around 600 °C are commonly associated with the onset of crystallization in sputtered $Ga_2O_3$ films. Consequently, a temperature range close to this value was selected to investigate the combined effect of discharge parameters, substrate temperature, and substrate type on film growth. The XRD patterns shown in Figure 5 reveal a pronounced temperature-dependent evolution of the crystallographic orientation of the deposited films.

For films deposited on quartz glass (Figure 5a), the sample prepared at 550 °C is XRD-amorphous and exhibits only a broad diffuse halo. Crystallization occurs only at temperatures above 600 °C, where diffraction peaks emerge and a pronounced preferred orientation develops. The (400) reflection becomes dominant, and the texture further strengthens with increasing temperature. Weak reflections from closely related planes, namely (401) and (801), are also observed in the diffractograms.

In contrast, films deposited on Si (100) (Figure 5b) exhibit an earlier onset of crystallization. Already at 600 °C, well-defined diffraction peaks are observed, indicating crystalline growth. With further increase in substrate temperature, significant changes in preferred orientation occur. At 660 °C, the (−201) reflection becomes dominant, with its higher-order reflections (−402) and (−603) also being clearly pronounced. At the same time, additional reflections such as (400), (402), and (603) remain present, indicating a strongly textured but not single-oriented structure.

For silicon substrates, the $Ga_2O_3$ films exhibit a mixed crystallographic orientation, with diffraction peaks corresponding to both the (200) and (−201) planes. The high intensity and narrow linewidth of the (−201) reflection at 660 °C indicate improved crystalline ordering and higher phase purity of the β-$Ga_2O_3$ phase. Crystallization proceeds predominantly via volume-driven nucleation rather than substrate-guided growth. These observations indicate that, on non-epitaxial substrates such as silicon and quartz glass, crystallization is governed primarily by bulk growth mechanisms, resulting in polycrystalline films with limited structural coherence. This behavior highlights the intrinsic limitations of non-epitaxial substrates for achieving highly oriented $Ga_2O_3$ films and motivates the use of epitaxial substrates, such as sapphire, where the crystallographic orientation can be imposed by the substrate (Figure 6).

The reflections from the sapphire substrate are denoted by asterisks (*). Minor peaks observed at $2\theta \approx 37.6°$ and 80° are attributed to Cu Kβ reflections originating from residual $K_\beta$ radiation of the x-ray source. The XRD data obtained on Si (100) and quartz glass are included for comparison. The corresponding diffractograms exhibit a similar overall shape to those shown in Figure 5 at 600 °C, although the samples were deposited at 585 °C.

FIG. 5. XRD patterns of films deposited at different substrate temperatures on quartz glass (a) and Si (100) (b). Black vertical lines indicate selected prominent reflections of the β-$Ga_2O_3$ powder standard (PDF 01-087-1901), labeled by their corresponding (hkl) indices.

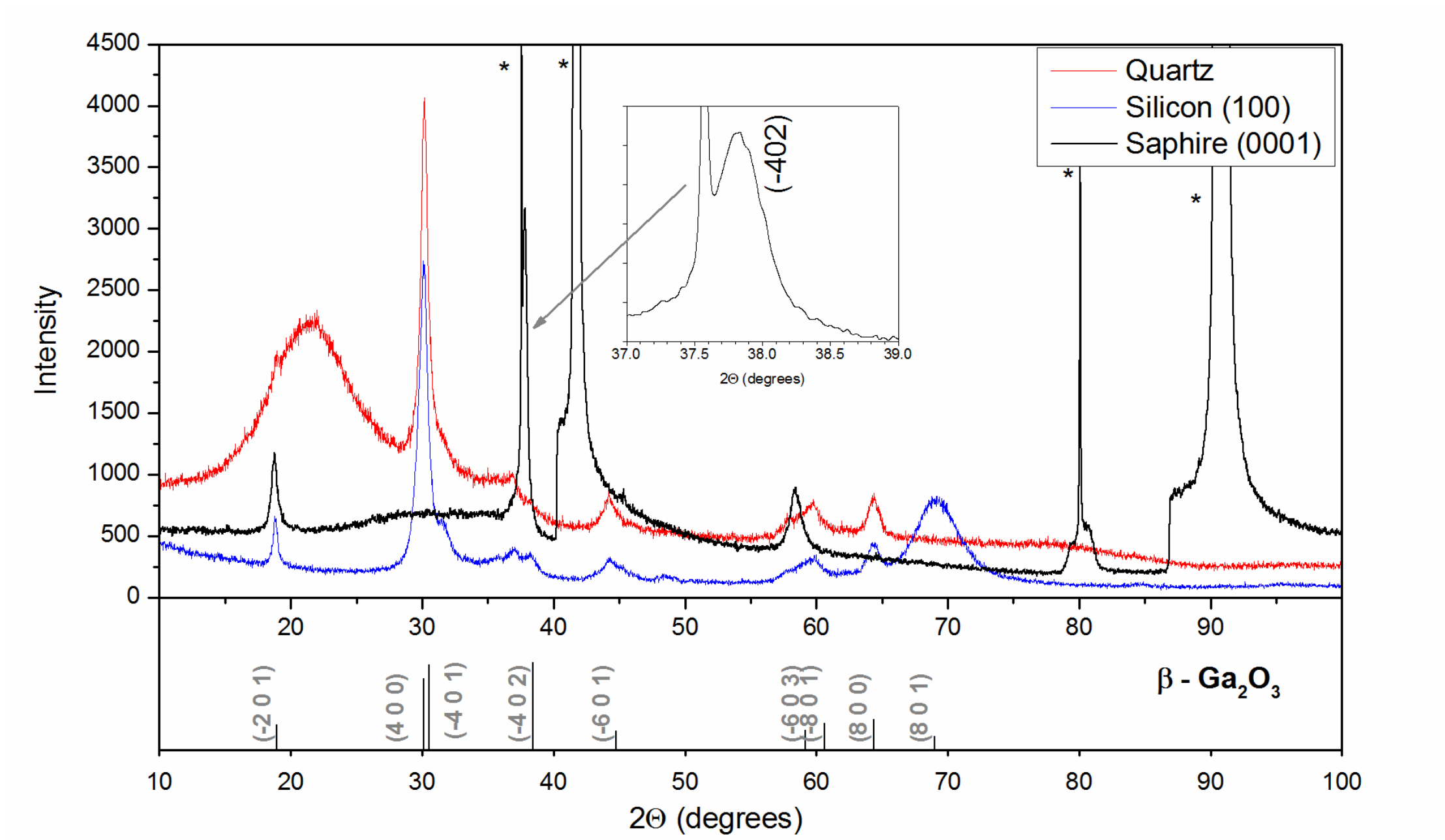

FIG. 6. XRD patterns of $Ga_2O_3$ films deposited on Si, $SiO_2$, and sapphire substrates under identical deposition conditions: substrate temperature 585 °C, discharge power 100 W, and $t_{off}$ = 85 µs. Grey vertical lines indicate selected prominent reflections of the β-$Ga_2O_3$ powder standard (PDF 01-087-1901), labeled by their corresponding (hkl) indices.

### C. *Growth and electrical properties of $Ga_2O_3$ films on sapphire glass*

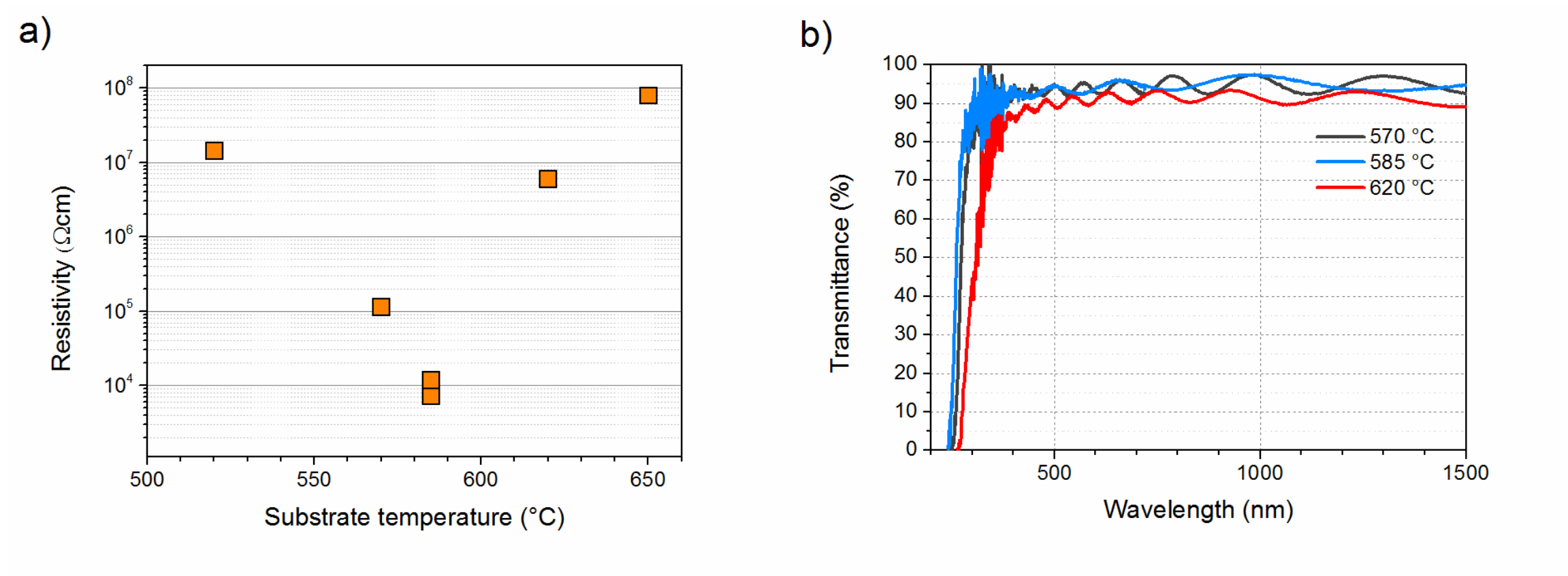

FIG. 7. Resistivity (a) and transmittance spectra (b) of $Ga_2O_3$ films deposited at different substrate temperatures.

In contrast to films deposited on non-epitaxial substrates, $Ga_2O_3$ films grown on sapphire show a pronounced preferential orientation along the (−201) crystallographic direction, which is commonly associated with epitaxial or epitaxial-like growth of β-$Ga_2O_3$ on sapphire substrates. This behavior indicates that the sapphire substrate provides a structurally favorable interface that fundamentally alters the growth mechanism compared to silicon and glass. The improved structural quality of the films deposited on sapphire is further evidenced by the narrowing of the XRD peaks and the appearance of well-defined higher-order XRD reflections. These observations suggest that the films exhibit enhanced crystallographic coherence and reduced phase disorder, indicating a clear enhancement in comparison to films deposited on non-epitaxial substrates under otherwise comparable deposition conditions. Consequently, $Ga_2O_3$ films grown on sapphire manifest a continuous and structurally stable morphology, enabling reliable assessment of their electrical properties (Figure 7a), while maintaining high optical transmittance in the visible range (Figure 7b). The optical band gap $E_g$ was estimated using Tauc plots, showing values of 4.9 eV and 5.0 eV for films deposited at 570 °C and 585 °C, respectively, followed by a decrease to 4.7 eV for the sample grown at 620 °C. This critical step is pivotal in evaluating the suitability of magnetron sputtering for the fabrication of $Ga_2O_3$ films intended for electronic applications. Resistivity is often influenced by the presence of poor crystallinity and excessive defect densities in sputtered films grown on non-epitaxial substrates.

Figure 7a shows the electrical resistivity of $Ga_2O_3$ films deposited on sapphire as a function of the substrate temperature during deposition. The electrical resistivity exhibits a pronounced dependence on the substrate temperature. For comparison, the resistivity of high-quality β-$Ga_2O_3$ single crystals can be estimated using typical experimentally reported values for unintentionally doped material. Assuming an electron concentration in the range of 2–4 × $10^{16}$ $cm^{-3}$ and an electron mobility of approximately 150 $cm^2$ $V^{-1}$ $s^{-1}$, the corresponding resistivity is expected to fall within the range of 1–2 Ω·cm[20]. These values represent an ideal benchmark for compact, defect-limited monocrystalline material and provide a useful reference for evaluating the electrical performance of sputtered $Ga_2O_3$ films prepared in this work.

With increasing temperature, the resistivity decreases and reaches a minimum at temperatures slightly below 600 °C. However, a further increase in the deposition temperature leads to a sharp rise in resistivity, despite the continued improvement in crystallinity observed in XRD measurements. This non-monotonic behavior suggests that factors other than crystallinity, such as defect formation or changes in carrier concentration, may play a significant role in determining the electrical performance.

XRD patterns of $Ga_2O_3$ films deposited on sapphire at different deposition temperatures, shown in Figure 8, demonstrate a strong dependence of crystallographic orientation on substrate temperature. The preferential orientation of β-$Ga_2O_3$ along the (−201) plane becomes clearly apparent at a deposition temperature of 585 °C and reaches its maximum at temperatures exceeding 600 °C. With increasing temperature, the intensity of the corresponding XRD peak maxima increases and higher-order reflections emerge, indicating improved crystallinity of the $Ga_2O_3$ films. Peaks marked with an asterisk correspond to diffraction from the sapphire substrate.

Although XRD analysis reveals a pronounced improvement in crystallographic order at elevated deposition temperatures (Figure 8), electron microscopy simultaneously indicates a loss of film compactness, likely associated with the formation of a nanocrystalline microstructure. This structural transition is clearly visible in the cross-sectional SEM micrographs (Figs. 9a–9c)

FIG. 8. XRD patterns of the $Ga_2O_3$ films deposited at temperatures from 520 to 620 °C.

The surface morphology of the film deposited at 520 °C (Fig. 9d) exhibits clearly distinguishable grain boundaries. Such grain boundaries are commonly associated with enhanced charge carrier scattering and limit carrier transport across the film, which can contribute to increased electrical resistivity.

Increasing the substrate temperature to 570 °C leads to grain coalescence, as shown in Fig. 9e, resulting in less pronounced grain boundaries. This microstructural evolution is consistent with the improved electrical conductivity observed at this temperature. Nevertheless, elevated deposition temperatures may also induce additional changes in the material that are not directly accessible by SEM analysis alone.

At a substrate temperature of 620 °C, the microstructure shows pronounced changes (Fig. 9f). The surface morphology reveals a finer and less uniform surface structure, which may adversely affect charge transport, even though individual grains exhibit improved crystallographic quality according to the XRD patterns. These microstructural changes are therefore likely related to the abrupt increase in electrical resistivity at higher deposition temperatures.

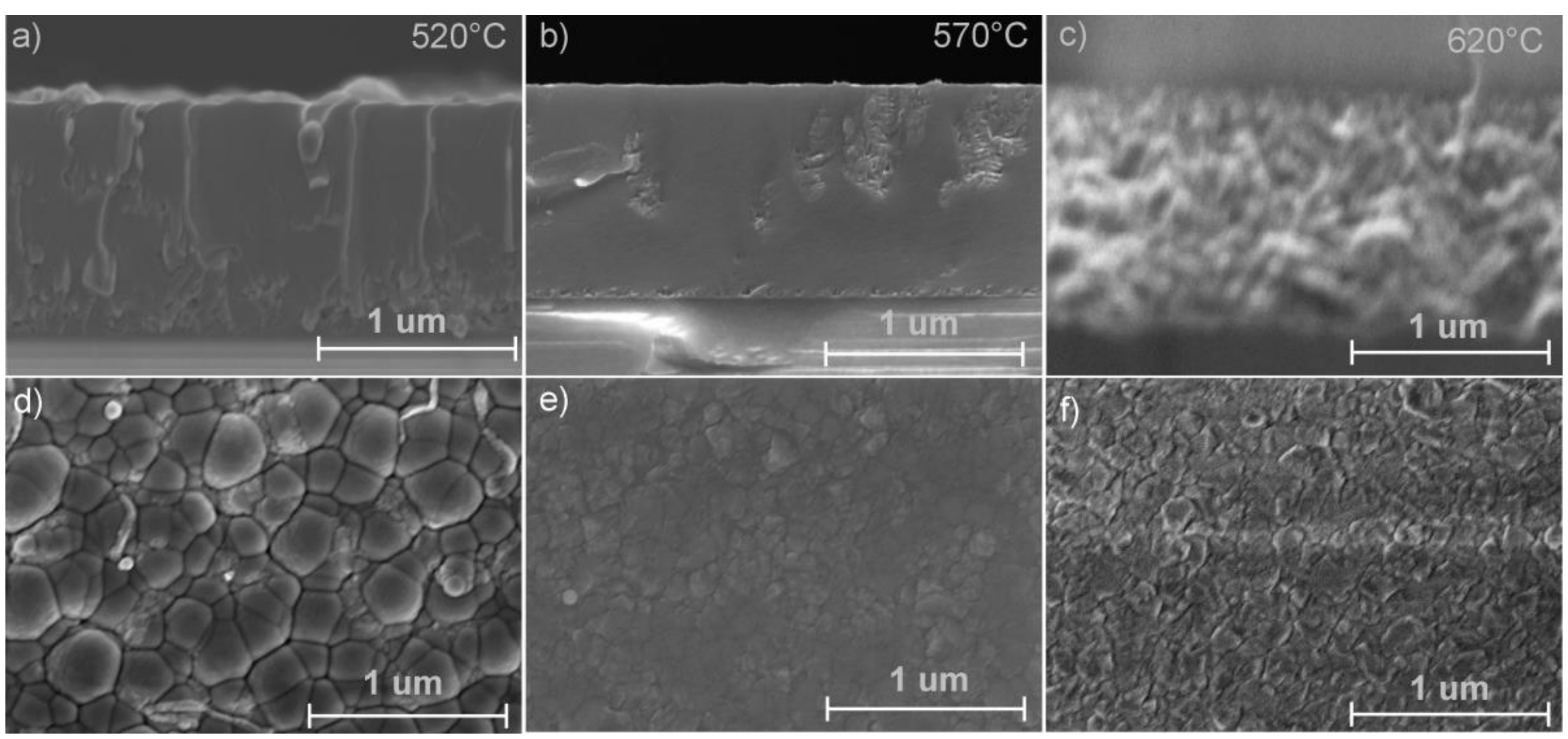

FIG. 9. Cross-sectional views of $Ga_2O_3$ films deposited on sapphire at substrate temperatures of (a) 520 °C, (b) 570 °C, and (c) 620 °C. The corresponding surface morphology is shown in panels (d), (e), and (f), respectively.

The observed behavior highlights a balance between crystallographic order and microstructural integrity in sputtered $Ga_2O_3$ films grown on sapphire. While sapphire enables strongly oriented growth and suppresses the formation of secondary phases, excessive deposition temperatures may promote microstructural rearrangement processes that limit the achievable electrical performance of $Ga_2O_3$ films.

In summary, the investigated films contain crystalline grains, as evidenced by XRD analysis. However, the films are not electrically homogeneous, as intergranular regions with reduced electrical conductivity may be present between the grains. The findings indicate that the most favorable deposition temperature is situated beneath the threshold of maximum crystallinity, where the crystallinity as determined by XRD is less optimal, yet this structure remains consistent throughout the film. It is evident that, at this temperature, the film undergoes more continuous and uniform growth, thereby attaining sufficient structural order. In such conditions, electrical resistance reaches its minimum, thereby emphasising the importance of the balance between crystallographic arrangement and microstructural integrity.

## IV. Conclusions

In summary, this study shows that the structural and electrical properties of sputtered $Ga_2O_3$ films can be influenced by the complex interactions between discharge conditions, deposition temperature, and substrate type. Although magnetron sputtering is not commonly used for epitaxial growth, sapphire glass substrates enable the formation of highly oriented compact $\beta$-$Ga_2O_3$ films under the investigated conditions. Therefore, the magnetron sputtering process itself may be a viable method for creating high-quality $Ga_2O_3$ films and requires further effort to explore the possibilities. However, the results show that optimal electrical performance is achieved at deposition temperatures below the crystallization threshold, where film growth remains compact despite incomplete crystallographic ordering. At higher temperatures, increased crystallinity is accompanied by microstructure degradation, leading to increased resistance. These findings define both the potential and the inherent limitations of magnetron sputtering for the preparation of electronically functional $Ga_2O_3$ films.

## DATA AVAILABILITY

The data that support the findings of this study are available from the corresponding author upon reasonable request.

## ACKNOWLEDGMENTS

*This work was supported by the project QM4ST, funded as project No. CZ.02.01.01/00/22_008/0004572 by P JAC, call Excellent Research.*

## AUTHOR DECLARATIONS

**Conflicts of Interest**

The authors have no conflicts to disclose